\documentclass[%
 reprint,
 superscriptaddress,
 showpacs, 
 amsmath,amssymb,
 aps,
 prl,
 floatfix,
]{revtex4-2}

\usepackage[pdftex]{graphicx}% Include figure files
    \graphicspath{{figures/}}
    \DeclareGraphicsExtensions{.pdf,.png}
\usepackage{dcolumn}% Align table columns on decimal point
\usepackage{bm}% bold math
\usepackage{color}
\usepackage{lineno} %  line numbers
\usepackage{multirow}
\usepackage[table]{xcolor}
\begin{document}

\title{ Strongly coupled Yukawa plasma layer in a harmonic trap}

\author{Hong Pan}
\affiliation{Department of Physics, Boston College, Chestnut Hill, Massachusetts, 02467, USA}

\author{Gabor J. Kalman}
\affiliation{Department of Physics, Boston College, Chestnut Hill, Massachusetts, 02467, USA}

\author{Peter Hartmann}
\affiliation{Institute for Solid State Physics and Optics, Wigner Research Centre for Physics, P.O.Box. 49, H-1525 Budapest, Hungary}

\date{\today}

\begin{abstract}
Observations made in dusty plasma experiments suggest that an ensemble of electrically charged solid particles, confined in an elongated trap, develops structural inhomogeneities. With narrowing the trap the particles tend to form layers oriented parallel with the trap walls. In this work we present theoretical and numerical results on the structure of three-dimensional many-particle systems with screened Coulomb (Yukawa) inter-particle interaction in the strongly coupled liquid phase, confined in one-dimensional harmonic trap, forming quasi-2D configurations. Particle density profiles are calculated by means of the hypernetted chain approximation (HNC), showing clear signs of layer formation. The mechanism behind the formation of layer structure is discussed and a method to predict the number of layers is presented. Molecular dynamics (MD) simulations provide validation of the theoretical results and detailed microscopic insights.

% When the strongly couple dusty plasma is confined in a trap, it shows inhomogeneity. In this paper, we firstly study the liquid density profile along the trapping direction by HNC approximation, the density profile has layer formation. Then we study the mechanism behind layer structure. We also design a method to predict the number of layers. All the analysis is based on the MD simulation.
\end{abstract}

\pacs{52.27.Gr, 52.27.Lw}
\maketitle

Strongly coupled charged particle systems (e.g.~plasmas), where the inter-particle interaction energy exceeds the thermal kinetic energy of the constituent particles, can often be approximated by the one component plasma (OCP) model~\cite{Baus1980}. In cases when the dynamics of one dominant particle species decouples from that of the other components, it might be sufficient to provide a detailed description only for the former, while approximating the contribution of rest with a continuous neutralizing background. In the case of warm dense matter, for instance, the ions can follow classical particle trajectories, while the electrons experience quantum degeneracy and realize a homogeneous background~\cite{Clerouin2016,Wang2020}. In the case of dusty plasmas it is the very different time scales and charge states that decouple the dust dynamics (with characteristic times in the order of 10~ms and $10^4$ elementary charges per dust) from the microscopic interactions with the electrons and ions present in the gas discharge (with typical times in the ns to $\mu$s range and unit charges). In both cases, only one of the plasma components (the ions in warm dens matter, and the dust particles in dusty plasmas) needs to be traced explicitly, the contribution of the background is reflected in the particular shape of the inter-particle interaction. If the polarizability of the isotropic neutralizing background is taken into account, the system can be approximated by the Yukawa one component plasma (YOCP) model. In this case the electrostatic pairwise interaction between the particles becomes screened, which screening can be approximated by the Dedye-Hückel mechanism resulting in an exponential decay superimposed to the bare Coulomb potential.

% One component plasma (OCP) is the simplest model to study the plasma. When the ions are screening by the background charges, the particle interaction becomes Yukawa form, which is called Debye screening. Dusty plasma and colloids are good examples for Yukawa one component plasma (YOCP). 

Since the pioneering experiments published in 1994~\cite{Chu94,Thomas94,Melzer94} laboratory dusty plasmas have been extensively used to gain insight into the microscopic details of macroscopic and collective phenomena. In large area radio frequency (RF) discharges, using micrometer sized monodisperse spherical dust particles it is easy to form a single layer of highly charged dust grains. This ensemble can form large (consisting of tens of thousands of dust grains) structures with the particles ordered in triangular lattices. Simply by changing discharge conditions (e.g.~the RF power) or by other forms of external energy coupling (e.g.~laser heating~\citep{Nosenko2006} or DC pulsing~\citep{Donko2017}) the system can undergo a solid to liquid transition and stabilize in the strongly coupled liquid state. However, the apparent single layer structure does not mean that the system is strictly two-dimensional. The vertical confinement is defined by the interplay of gravity and the electric field in the RF sheath, ultimately forming a potential well experienced by each dust particle. Both the vertical equilibrium position and the effective trap frequency depend on the charge-to-mass ($q/m$) ratio of the dust grains and the discharge conditions. At finite temperature the dust particles experience small amplitude vertical oscillations, forming quasi-2D configurations.

Early experimental investigations on quasi-2D system have been published by Lin I~\cite{Juan1998,Teng2003} observing anomalous diffusion, as well as layering and slow dynamics. Theoretical work on quasi-2D systems include studies of in-plane and out-of-plane polarized wave propagation and structural transitions into multi-layered configurations \cite{Bystrenko2003,Hartmann2005,Qiao2005,henning2007,Donko2009} building heavily on numerical simulations. An analytical approach targeting the particle density distribution in the trap would provide deeper insight into the double or triple layer formation, as observed in the experiments. A similar problem has been studied by Wrighton~\cite{Wrighton2009,Bruhn2011,wrighton2012} for spherical confinement comparing mean field and hypernetted chain calculations to numerical results. In Klumov’s work \cite{klumov}, crystallization of quasi-2D system under both harmonic and hard wall confinement was studied, in which the wetting phenomenon was observed in the hard wall case. Further theoretical studies on radially, as well as linearly confined liquids~\cite{vanderlick1989,yu2009,Matheus2014,Nyg_rd_2016,mansoori2014,henning2006} include the prediction of the radial density distribution, mechanisms of oscillatory density profiles and ensuing solvation forces, in both the weak and strong coupling regimes. 

In this work, we study the evolution of the density profile of quasi-2D Yukawa systems in a one-dimensional harmonic trap in the strongly coupled liquid state. The system is infinite in the $x$ and $y$ directions, where the particles can move freely. A harmonic trapping potential is applied in the $z$ direction. Let $n_{\rm s}$ denote the surface density of the quasi-2D layer projected onto the $x$-$y$ plane. In this case we define $a$ as the Wigner-Seitz radius in the projected plane as $a^2 = 1/(\pi n_{\rm s})$. 

The Yukawa interaction potential energy and the harmonic trap potential are, respectively
\begin{equation}  \label{eq:potential}
    \varphi(r)=q^2\frac{e^{-\kappa \,r/a}}{r},~~~~~{\rm and}~~~V(z)=\frac{m \omega_{\rm t}^2 z^2}{2},
\end{equation}
where $r$ is the three-dimensional inter-particle distance, $\kappa$ is the dimensionless Yukawa screening parameter, $\omega_{\rm t}$ is the trap frequency, $q$ and $m$ are the electric charge and mass, equal for all particles.

The strength of the electrostatic coupling can be characterized with the Coulomb coupling parameter, nominally expressing the ratio of the potential to kinetic energies per particle, and is defined as
\begin{equation} \label{eq:Gamma}
    \Gamma=\beta q^2/a,
\end{equation}
where the thermodynamic $\beta = 1/(k_{\rm B}T)$. For Yukawa systems an effective coupling $\Gamma_{\rm eff}(\kappa)<\Gamma$~\cite{Hartmann2005,Ott2014} can be introduced, that depends on the Yukawa screening parameter a does more accurately characterize the state of a particular system. The strongly coupled domain is defined by $\Gamma \gg 1$.

Using the nominal 2D plasma frequency $\omega_{\rm p}^2=2\pi q^2 n_s/(ma)$ to define the frequency and time units we introduce the dimensionless parameter $t$ characterizing the strength of the trapping potential, as
\begin{equation} \label{eq:trap}
    t^2=\frac{\omega_{\rm t}^2}{\omega_{\rm p}^2}=\frac{m \omega_{\rm t}^2 a}{2 \pi q^2 n_{\rm s}} = \frac{m \omega_{\rm t}^2 a^3}{2q^2}.
\end{equation}

Within the frame of the quasi-2D YOCP model the behavior of a system is fully determined by the three dimensionless parameters $\Gamma$, $\kappa$ and $t$.

Both the equilibrium properties and the dynamics of the system have been studied by molecular dynamics (MD) simulation and by theoretical analysis. Here we report on the equilibrium studies. Our MD simulations trance the trajectories of 10\,000 particles in an external trapping potential defined by $V(z)$ and periodic boundary conditions in $x$ and $y$ in a cubic simulation domain. Initial positions are assigned to the particles based on a simple initial barometric estimate. Inter-particle forces are summed for all particles within a radius of $R \approx 44 a$ for each particle. Before conducting any measurements the system is given enough time (approximately of 4\,000 plasma oscillation cycles) to reach equilibrium during the initial thermalization phase using the velocity back-scaling thermostat. This is verified by observing the temperature stability after the thermaization is turned off. During the measurements data is collected and averaged over approx.~2\,000 plasma oscillation cycles (100\,000 time-steps). 

In order to obtain a theoretical description of the density distribution within the trap we invoke the density functional theory (DFT) \cite{HansenBook,Evans}. By minimizing the grand potential, a general, but quite complex expression for $n(r)$, the 3D number density has been reported by Evans~\cite{Evans} (eq.~26 in chapter~3), see also the Appendix of~\cite{Wrighton2009} in terms of $c\left(r,r';[n(r)]\right)$, the direct correlation function (DCF) of the system. The notation emphasizes that the DCF is a unique, albeit unknown functional of the density profile. $c(r,r')$ is also connected with $h(r,r')$, the pair correlation function through the Ornstein-Zernike equation (OZ).
\begin{equation} \label{eq:hrr}
    h(r,r')=c(r,r')+\int c(r,r'') n(r'') h(r'',r')~{\rm d}r''
\end{equation}
The latter is related to the pair distribution function through $g(r,r')=h(r,r')+1$.

The density profile is determined by the self-consistency relation \cite{HansenBook}
\begin{equation} \label{eq:nr}
    n(r) \propto \exp\left[ -\beta V(r) + \int n(r') c_0(r-r';n)~{\rm d}r' \right],
\end{equation}
where $c_0$ is the DCF of a reference system with uniform density. The structure of eq.~(\ref{eq:nr}) tells us that $c(r,r')$ plays the role of the effective interaction potential in the system (note that $\varphi(r)$ does not appear explicitly in eq.~(\ref{eq:nr})),
\begin{equation} \label{eq:beta}
    -\beta \varphi_{\rm eff}(r-r')=c_0(r-r')
\end{equation}

In order to solve $c$ and $h$ simultaneously, the OZ relation provides the first equation, the second relation is derived by applying the hypernetted chain (HNC) approximation~\cite{Rowlinson_1965,HansenBook}. HNC has been successfully used in various problems relating to strongly coupled Coulomb and Yukawa systems~\cite{Ng1974}. The HNC approximation is based on the neglect of the so-called ``bridge'' (or irreducible, i.e.~not derivable from the combined operations of ``parallel connection'' and ``series connection'' of Mayer diagrams) diagrams: their contribution is assumed to be negligible in the case of long range potentials. As a result, one obtains the general basic relationship
\begin{equation} \label{eq:hnc1}
    g(r,r')=\exp\left[ -\beta\varphi(r-r')+h(r,r')-c(r,r') \right],
\end{equation}
which in combination with the OZ equation (\ref{eq:hrr}), provides a solution for $h(r,r')$ and $c(r,r')$. Restricting the solutions to homogeneous and isotropic functions, DCF satisfies the simpler variant of the OZ equation
\begin{equation}
\label{eq:uoz}
    h(r) = c(r) + {\bar n}\int c(|r-r'|)h(r')~{\rm d}r'.
\end{equation}
In equation (\ref{eq:hnc1}), when $r$ is large enough, $g(r) \rightarrow 1$, $h(r) \rightarrow 0$, the expression reduces to $-\beta \varphi_(r-r')=c(r-r')$. Using this asymptotic formula for the whole range of $r$, one arrives to the mean field (MF) approximation. Another self-consistent approach to calculating DCF for homogeneous fluids was described in~\cite{rickayzen1994}.

Since the system is uniform in the $x$-$y$ plane, the density is non-uniform only along the $z$ direction, the parametrization can be simplified to $n(r)=n(z)$, with the normalization condition $n_s=\int n(z)~{\rm d}z$. The integral in equation (\ref{eq:nr}) can be split into$z$ part and radial part separately. We can write eq.~(\ref{eq:nr}) as 
\begin{equation} \label{eq:nrp}
    n(z) \propto \exp\left[ -\beta V(z) +2\pi \iint n(z') c_0(z,z',\rho';n)\rho'~{\rm d}\rho'~{\rm d}z'\right].
\end{equation}

We can rewrite eq.~(\ref{eq:nrp}) in terms of the dimensionless quantities $\tilde{n}=na^2$, $\tilde{z}=z/a$, $\tilde{U}(z)=\frac{\beta}{\Gamma} U(z)$, etc. as
\begin{eqnarray} \label{eq:nz}
    \tilde{n}(\tilde{z}) &=& \tilde{n_s}\frac{\exp[-\Gamma \tilde{U}(\tilde{z})]}{\int_{-\infty}^{\infty} \exp[-\Gamma \tilde{U}(\tilde{z})]~{\rm d}\tilde{z}} \\
    \tilde{U}(\tilde{z}) &=& t^2\tilde{z}^2-\tilde{W}(\tilde{z}) \nonumber \\
    \tilde{W}(\tilde{z}) &=& \frac{2\pi}{\Gamma} \iint  \tilde{n}(\tilde{z}') c_0(\tilde{z},\tilde{z}',\tilde{\rho}';n)\tilde{\rho}'~{\rm d}\tilde{\rho}'~{\rm d}\tilde{z}'
    \nonumber
\end{eqnarray}

$\rho$ being the 2D distance between two projected particle positions in the $x$-$y$ plane. Then, provided that $c_0(r-r';n)$ is known one can obtain the density profile by the iterative solution of eq.~(\ref{eq:nz}).
In view of eq.~(\ref{eq:beta}) the simplest approximation for $c_0(z-z',\rho)$ is to ignore correlations and set $\varphi_{\rm eff}(r,r')=\varphi(r-r')$, i.e.
\begin{equation} \label{eq:c0}
    c_0(r-r')=-\beta \varphi(r-r')
\end{equation}
This is tantamount to a mean field approximation. Substituting (\ref{eq:c0}) into (\ref{eq:nz}), (\ref{eq:nz}) reduces to
\begin{eqnarray} \label{eq:Uz}
    &U(z) = t^2 z^2 + 2 \iint n(z')\frac{\exp[-\kappa d(z,z',\rho')]}{d(z,z',\rho')}\rho'~{\rm d}\rho'~{\rm d}z' \nonumber \\
    &d(z,z',\rho') = \sqrt{(z-z')^2+\rho'^2}.
\end{eqnarray}

\begin{figure}[htb]
\begin{center}
\includegraphics[width=\columnwidth]{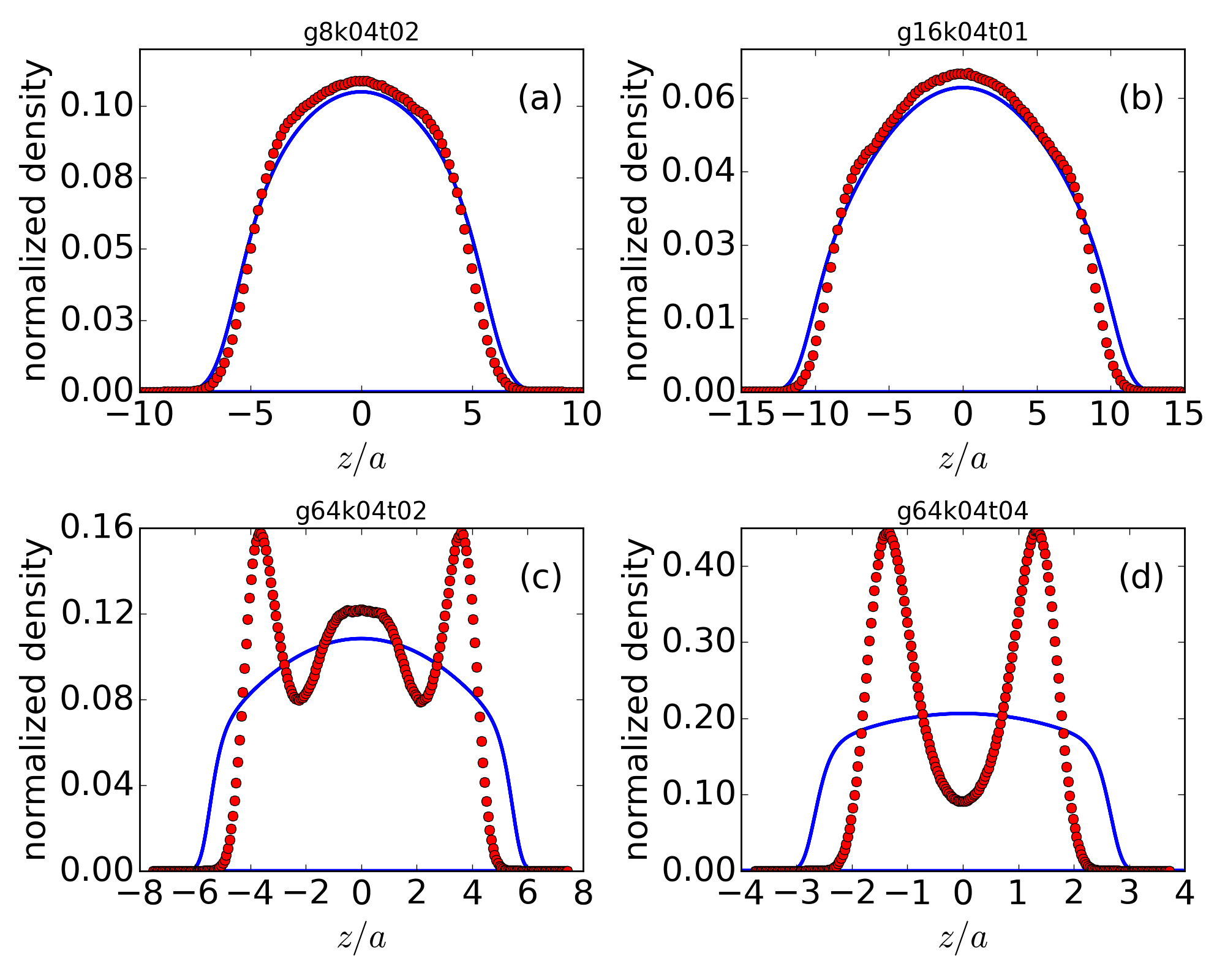}
\caption{Density profile comparison between the MD simulation and the MF approximation (continuous line) for different $\Gamma$ and t values. (a) $\Gamma=8$, $t=0.2$,  (b) $\Gamma=16$, $t=0.1$,  (c) $\Gamma=64$, $t=0.2$,  (d) $\Gamma=64$, $t=0.4$}
\label{fig:MD1}
\end{center}
\end{figure}

In the sequel we drop the symbol for the dimensionless quantities. It should be kept in mind that all the length variables are in the unit of the Wigner-Seitz radius $a$. In the following, for simplicity, we only discuss the results for $\kappa=0.4$. The results of the MF calculation and their comparison with the results of the MD simulation are given in Fig.~\ref{fig:MD1}. If there is no particle-particle interaction, then $\Gamma \rightarrow 0$, the profile is Gaussian $n(z)\propto \exp[-\Gamma t^2 z^2]$, mapping the Maxwell distribution of non-interacting particles. Proceeding to higher $\Gamma$ values, we expect the MF method to gradually fail to reproduce the numerical data as it is only valid for weak coupling (low $\Gamma$, i.e.~low density or high temperature). Inspecting the MF profiles in Fig.~\ref{fig:MD1}, covering moderate and strong $\Gamma$ values, we observe that those reasonably match the MD results at low $\Gamma$ values, while fail spectacularly at high $\Gamma$-s. In particular MF does not provide the non-monotonic behavior related to layer formation. In fact, it has been proved analytically that the MF density profiles have to be monotonic on both sides of the maximum~\cite{Hongthesis}, and thus MF is unable to predict the formation of multiple layers.

\begin{figure}[htb]
\begin{center}
\includegraphics[width=\columnwidth]{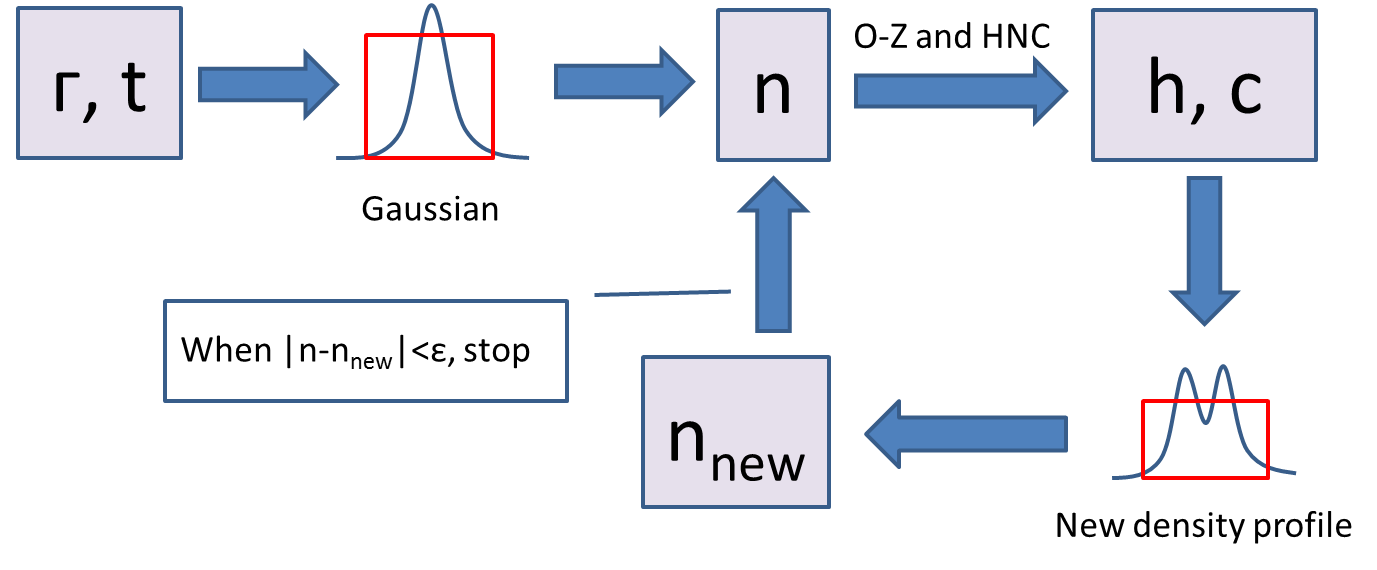}
\caption{Numerical iteration loop for calculating density profile.}
\label{fig:HNC}
\end{center}
\end{figure}

Next we calculate the DCF from the HNC approximation, then derive the density profile. The $\bar n$ in eq.~(\ref{eq:uoz}) is chosen as the average number density of the plasma between the two layer boundaries. The boundary is defined arbitrarily as the points where the density value is one percent of the maximum density in the layer. This, of course requires the knowledge of the density, whose determination is the purpose of the calculation. All this lends itself to an iteration scheme. The resulting protocol for the numerical calculation is portrayed in Fig.~\ref{fig:HNC}. The algorithm is based on work published in~\cite{Springer1973}.

\begin{figure}[htb]
\begin{center}
\includegraphics[width=\columnwidth]{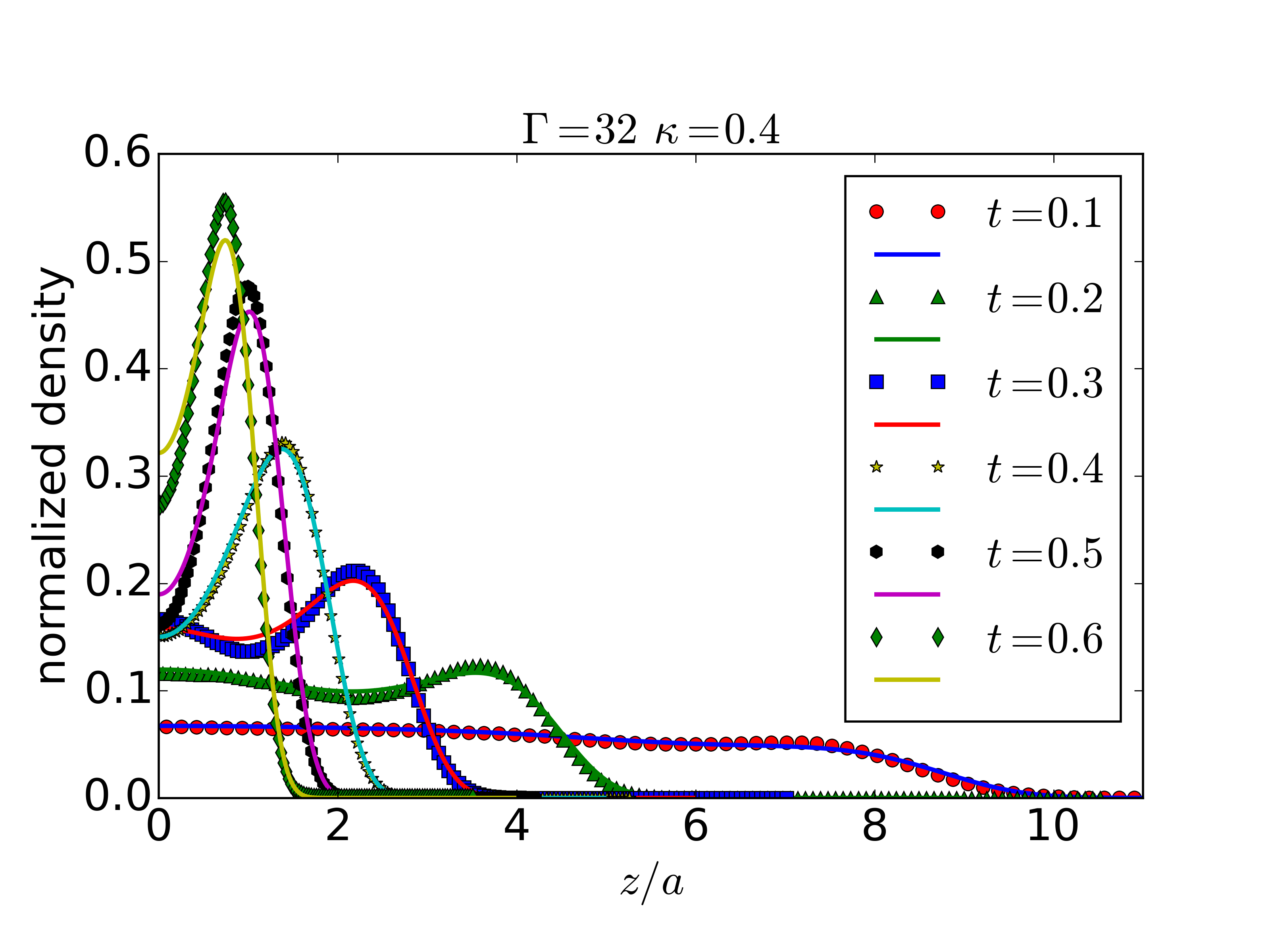}
\caption{Density profile comparison between MD simulation (continuous line) and the HNC approximation, for different trapping strengths. $\Gamma=32$}
\label{fig:MD2}
\end{center}
\end{figure}

\begin{figure}[htb]
\begin{center}
\includegraphics[width=\columnwidth]{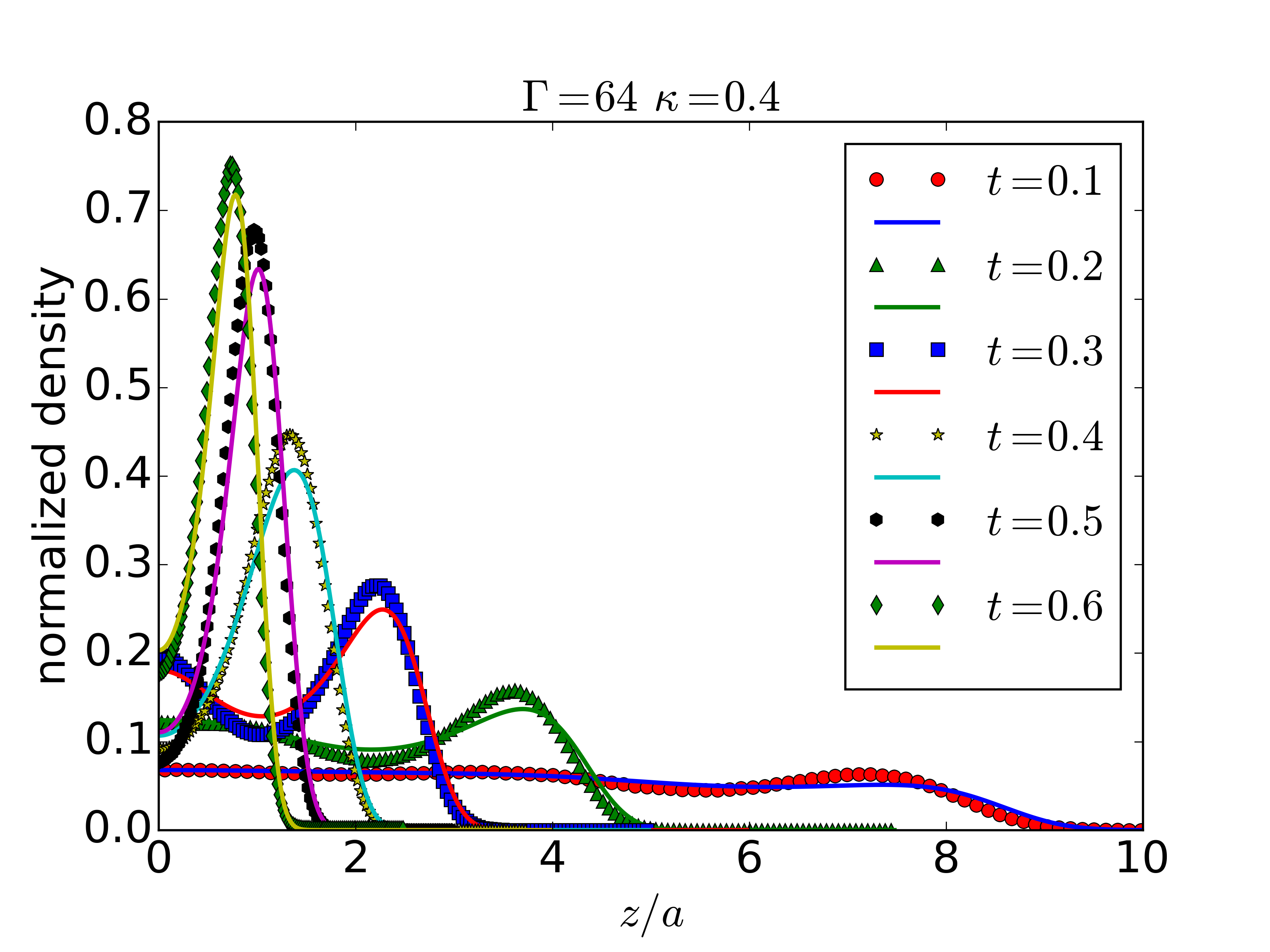}
\caption{Density profile comparison between MD simulation (continuous line) and the HNC approximation, for different trapping strengths. $\Gamma=64$}
\label{fig:MD2a}
\end{center}
\end{figure}

%We note that a by-product of the calculation is the determination of the pair correlation function   for a homogeneous 3D YOCP for a range of ${\bar \kappa}$, $\Gamma$ values. These may be compared with available MD data in the literature [...]. We will return to this point later below.

The major improvement of the HNC over the MF calculation is that it correctly reproduces the splitting of the system into multiple layers. Figs.~\ref{fig:MD2} and \ref{fig:MD2a} show comparisons with MD data at moderately high coupling values for cases when multi-peak profiles form. The similar multi-peak profile was reported in previous works \cite{snook1980,vanderlick1989,Nyg_rd_2016,mansoori2014,trond2014,ebner1980,kierlik1991}.

Generally, both $t$ and $\Gamma$ can affects the density profile formation, but the acting mechanisms are different. From Figs.~\ref{fig:MD2} and \ref{fig:MD2a} we can see that when $t$ decreases, the layer becomes wider, developing more peaks (layers) in the density profile. The coupling parameter $\Gamma$ only affect the density modulation amplitude. The relation between the trapping strength and the density profile was also studied in~\cite{ott2008,almarza2009}.

%Increasing the coupling  further (or relaxing the trapping strength) leads to the appearance of  triple, quadruple, etc. layer structures, as shown in Fig.~\ref{fig:TH1}, where the evolution from monotonic to layered structure can be observed.
%Fig.~\ref{fig:TH2} displays a sequence of different layer structures in the $t$ -- $\Gamma$ plane, as determined by MD simulations. 

\begin{figure}[htbp]
\begin{center}
\includegraphics[width=\columnwidth]{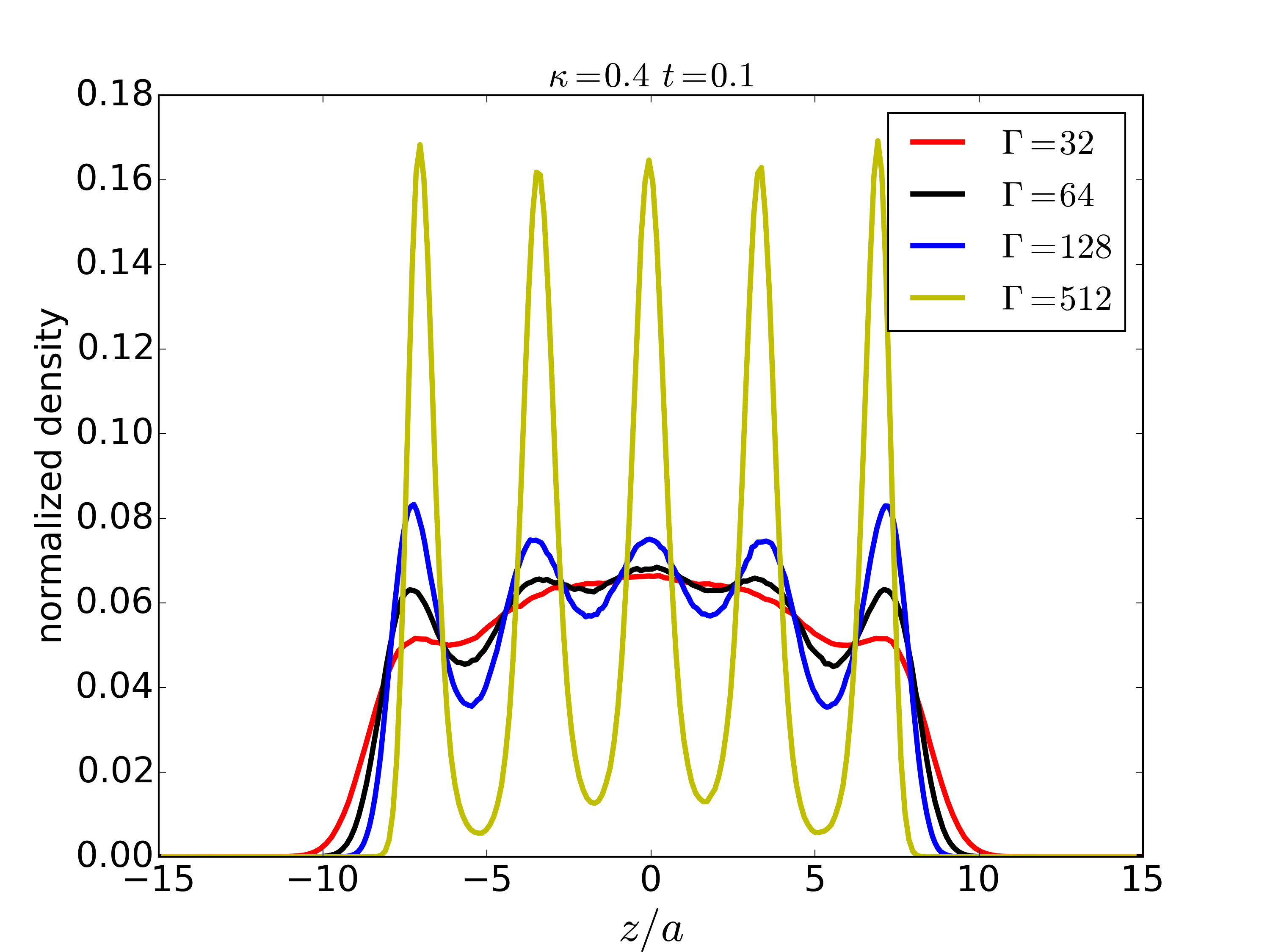}
\caption{Evolution of the density profiles with increasing $\Gamma$ values as determined by MD simulations. $t=0.1$}
\label{fig:MD3}
\end{center}
\end{figure}

\begin{figure}[htbp]
\begin{center}
\includegraphics[width=\columnwidth]{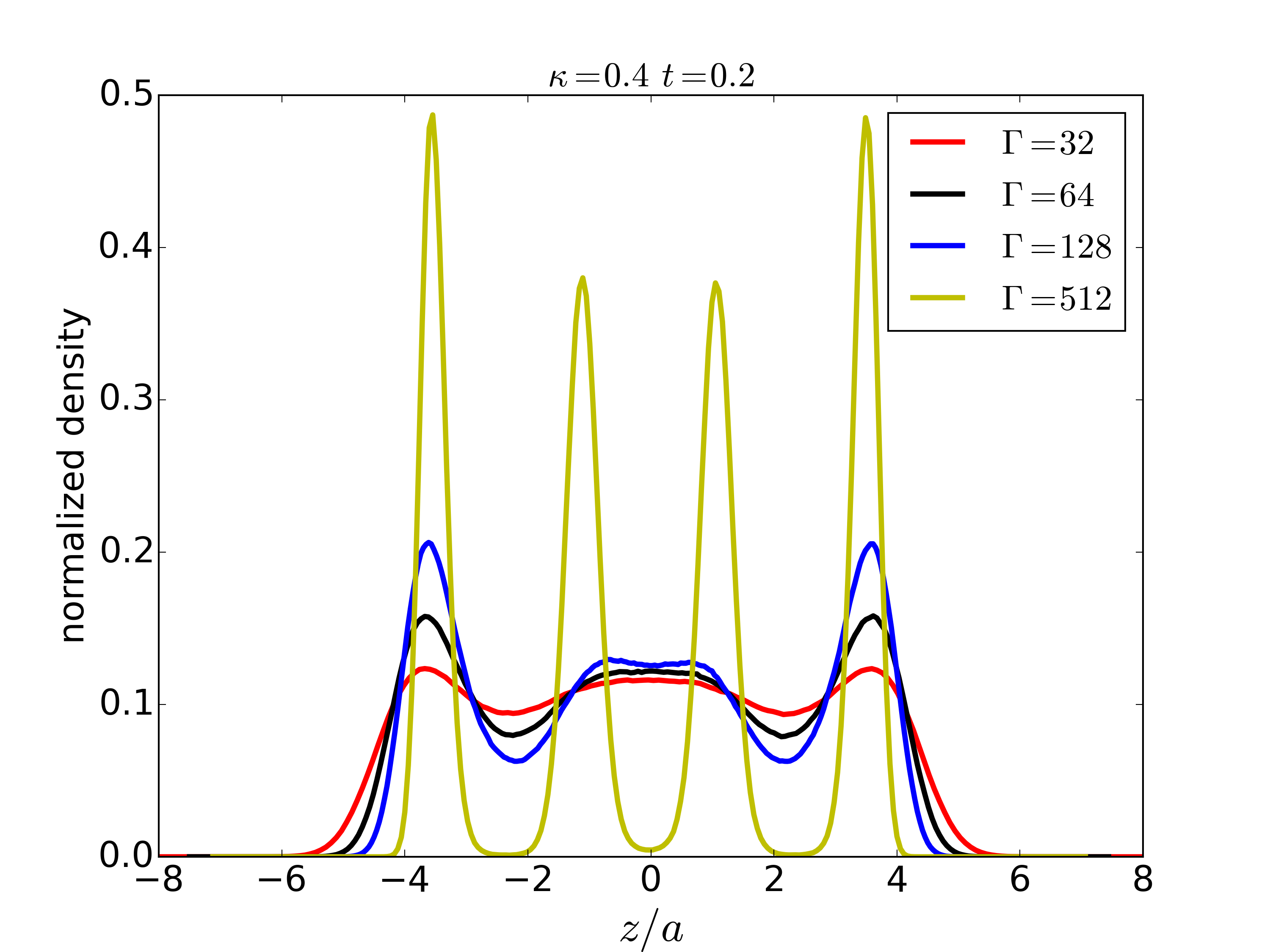}
\caption{Evolution of the density profiles with increasing $\Gamma$ values as determined by MD simulations. $t=0.2$}
\label{fig:MD3a}
\end{center}
\end{figure}

MD results in Figs.~\ref{fig:MD3} and \ref{fig:MD3a} show density profiles for a set of $\Gamma$ values. With increasing $\Gamma$, the system develops very sharp density peaks with deep minima between the peaks, a sign of the formation of crystal-like ordering (no exchange of particle between layers). As $\Gamma$ decreases, the particles experience larger vertical oscillation amplitudes, and the sides of neighboring peaks in the density profile fuse. The position and the number of the density peaks is mostly independent of the coupling, the distribution in the liquid state resembles that of the solid with lower amplitude of the density modulation.

%At the same $\Gamma$ values at which multiple layers form, the YOCP undergoes a liquid-solid phase transition and the layers crystallize.

\iffalse %%%%%%%%%%%%%%%%%%%%%%%%%%
\begin{figure}[htb]
\begin{center}
\includegraphics[width=\columnwidth]{Fig_TH1}
\caption{???}
\label{fig:TH1}
\end{center}
\end{figure}

\begin{figure}[htb]
\begin{center}
\includegraphics[width=\columnwidth]{Fig_TH2}
\caption{Table of normalized density profiles from MD simulation for   $2 < \Gamma <128$ and for $0.1< t <10$.   Note that the $x$-scales in the three boxes are different. AXES NEED LABELLING - GJK }
\label{fig:TH2}
\end{center}
\end{figure}
\fi %%%%%%%%%%%%%%%%%%%%%%%%%%%%%%%%%

%The phase transition boundary $\Gamma^*_2({\bar \kappa})$ for a 2D YOCP] and $\Gamma^*_3({\bar \kappa})$ for a 3D YOCP have been known for some time [HARTM, DUBIN], ...]. Here the critical coupling value depends also on $t$: Fig.~\ref{fig:DIFF} shows the $\Gamma^*(t,{\bar \kappa})$ behavior.  

%MORE - GJK

\begin{table*}[tb]
\centering
\begin{tabular}{|c|c|c|c|c|c|c|c|c|c|c|c|c|}
\hline
\multicolumn{2}{|c|}{ \multirow{2}*{} }& 3 &3&3&4&4&5&5&6&6&7& peak number\\
\cline{3-13}
\multicolumn{2}{|c|}{}&0.34&0.3&0.22&0.2&0.14&0.12&0.09&0.08&0.07&0.06&t\\
\hline
N&$\sqrt{N\pi}$&3.5&4.2&6&7&10&12&15&17.5&19&22&layer width\\
\hline
2&2.5066&3.5&4.2&6&7&10&12&15&17.5&19&22& \multirow{7}*{$\frac{\mbox{layer width}}{N-1}$}\\
\cline{1-12}
3&3.0700&\cellcolor{green}1.75&\cellcolor{green}2.1&\cellcolor{green}3&3.5&5&6&7.5&8.75&9.5&11&\\
\cline{1-12}
4&3.5449&1.1667&1.4&2&\cellcolor{green}2.3333&\cellcolor{green}3.3333&4&5&5.8333&6.3333&7.3333&\\
\cline{1-12}
5&3.9633&0.875&1.05&1.5&1.75&2.5&\cellcolor{green}3&\cellcolor{green}3.75&4.375&4.75&5.5&\\
\cline{1-12}
6&4.3416&0.7&0.84&1.2&1.4&2&2.4&3&\cellcolor{green}3.5&\cellcolor{green}3.8&4.4&\\
\cline{1-12}
7&4.6895&0.5833&0.7&1&1.1667&1.6667&2&2.5&2.9167&3.1667&\cellcolor{green}3.6667&\\
\cline{1-12}
8&5.0133&0.5&0.6&0.8571&1&1.4286&1.7143&2.1429&2.5&2.7143&3.1429&\\
\hline
\end{tabular}
\caption{Number of layers with different trapping strength, $\Gamma=512$, $\kappa=0.4$, the row with green color corresponds to the predicted $N$ value.}
\label{table:t1a}
\end{table*}

%The determination whether the system is in the liquid or solid phase has been done by following the time evolution of X-Y plane mean square displacement $\xi(t)$ of the plasma particles from a given original configuration. The $t \rightarrow 0$ behavior can be characterized by $\xi(t\rightarrow 0) \propto t^H$. The data show clear distinction between groups with $H = 1$ and $H \ll 1$, the former we identify as marking a solid and the latter a liquid phase. Some work about the dusty plasma diffusion can be found in \cite{Hou2009}.

% In the 3D lattice the distance $d$ between the layers should not exceed the 2D WS radius consistent with the layer density $n_l = n_s/L$. On the other hand, the total width w of the entire trapped plasma  can be estimated by assuming that it is  that of  the  enveloping interaction-free Gaussian, (cf. Fig.~\ref{fig:MD1}) $w\propto 1/(t\sqrt{\Gamma})$ and $d=w/(N-1)$. Thus as $t$ decreases, at a critical ($\Gamma$-dependent) value a new layer emerges. This is illustrated in the graphical construction of Fig.~\ref{fig:MD3}, which provides an excellent prediction for $N$ as a function of $t$. 

Based on the idea that a multiple layers may be regarded as a slab extracted from a 3D lattice~\cite{Totsuji1997,Totsuji1996,Oguz2009}, one may determine the number of layers by a simple algorithm. Assuming there are $N$ layers, for simplicity a planar square lattice is associated to each layer. In this case the unit square side length is $x_N=\sqrt{\pi N}$. The inter-layer distance is $d_N=\frac{w}{N-1}$, where $w$ is the distance between the two outermost peaks. Comparing $x_N$ and $d_N$, the lowest value of $N$ at which $x_N>d_N$ is the prediction for the number of layers (see Table~\ref{table:t1a}). In~\citep{Totsuji1997} the total energy of the system with different number of layers, including the particle interaction and the trapping energy is calculated. Despite operating with different physical quantities, the principles of the two methods are equivalent. In~\cite{ricci2007}, the shell structure of the density profile in a confined colloidal particle system was studied by Monte Carlo simulation.

%\onecolumngrid
\iffalse
\begin{figure}[htbp]
\begin{center}
\includegraphics[width=\columnwidth]{Hong paper/g512k04.png}
\caption{Number of layers with different trapping strength, $\Gamma=512$, $\kappa=0.4$, the row with green color corresponds to the predicted $N$ value.}
\label{fig:tblay1}
\end{center}
\end{figure}
\fi
%\twocolumngrid
%Finally, we return to the question of how the nHNC generated $g(r)$  correlation functions agree with either the MD generated YOCP $g$-s, or with the MD generated $g(\rho,z)$ correlation functions. 

%THIS HAS TO BE CLARIFIED - GJK

%Lastly, what's the upper bound of gamma can we apply in HNC approximation? In our test, it is about 200. But at higher gamma, the discrepancy is larger.

\begin{figure}[htbp]
\begin{center}
\includegraphics[width=\columnwidth]{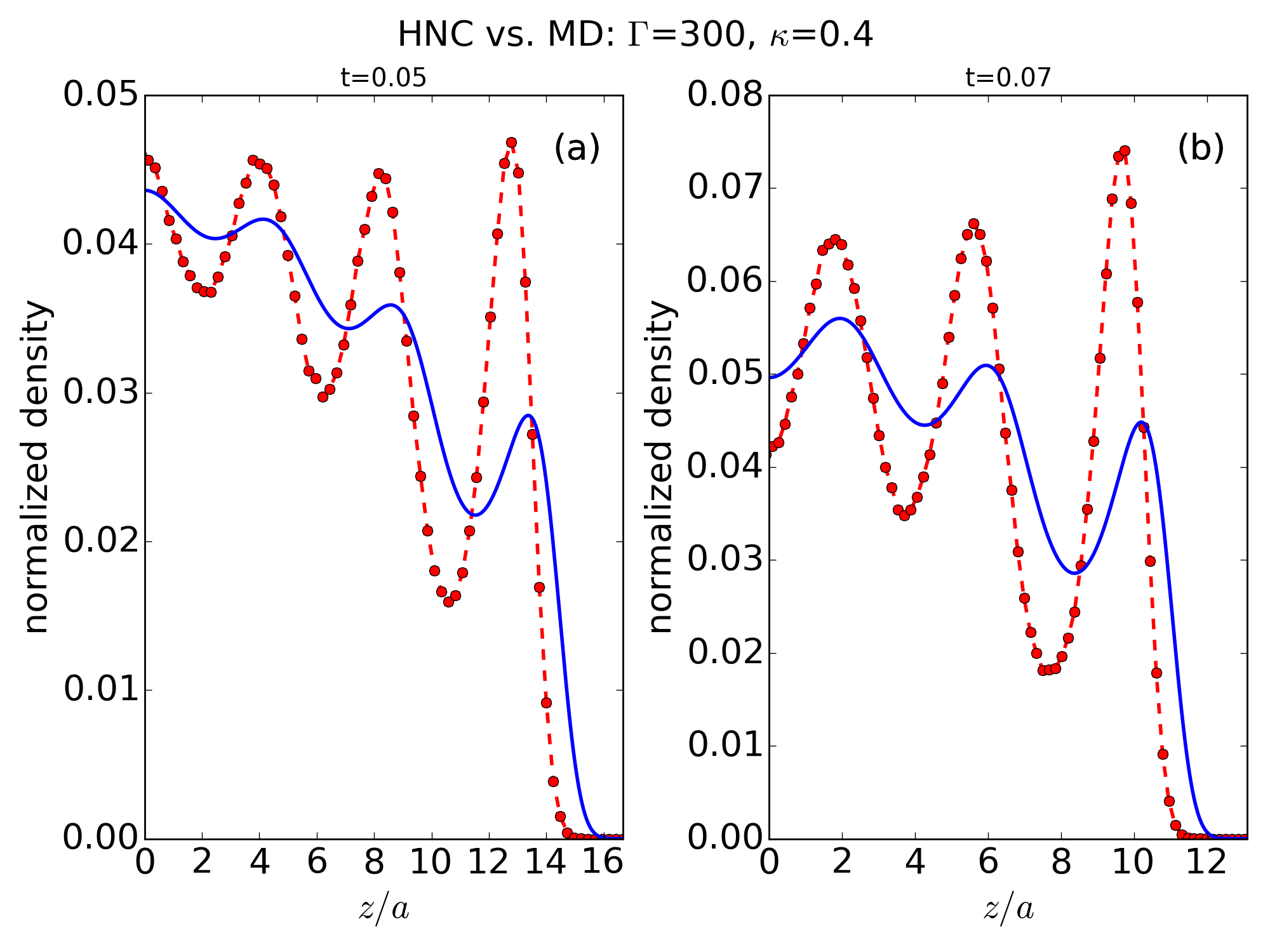}
\caption{HNC (lines) and MD (symbols) density profile comparison at high $\Gamma$. $\Gamma=300$. (a) $t=0.05$, (b) $t=0.07$. }
\label{fig:MD4}
\end{center}
\end{figure}

A similar configuration was studied in~\cite{Matheus2014} applying both the MF and HNC formalism, however, as those results are restricted to the weak coupling regime no layer formation was reported. In that work the correlation energy, based on local density approximation (LDA), and with this the excess chemical potential was calculated. In anther study~\cite{dutta2016} the density profile of a polyelectrolyte system in a harmonic trap was calculated. A similar shell structure formation was observed. In~\cite{gu2012} a similar DFT method was applied on hard-core Yukawa dusty plasma in a spherically harmonic trap. In~\cite{yiping2004} a study on van der Waals fluids with different confining potentials was presented including the modulation of the density profiles.

In our calculations, relying on the HNC approximation, the stability of the solution scheme appears to be limited to $\Gamma\le 300$ (for $\kappa=0.4$), at higher couplings no converged solution could be found. For this reason comparing the HNC and MD result at the strongest coupling is performed at $\Gamma=300$. Fig.~\ref{fig:MD4} shows that even though the HNC result underestimates the density modulation amplitude, it predicts correctly the number and position of the peaks.

In conclusion, in this paper, we have analyzed the density profile of quasi-2D Yukawa plasma in harmonic trap in the strongly coupled liquid regime. We utilize the HNC approximation to calculate the density profile, which matches MD simulation result very well at low to moderate couplings. At stronger coupling, near the solidification the HNC calculations underestimate the density modulation amplitudes. We have presented a method to predict the number of layers in the solid phase based on the comparison between the in-plane inter-particle distance and the inter-layer distance. We confirm the validity of this prediction in the liquid state as well. 

An outlook into future research directions include: (1) theoretical models beyond the standard HNC approximation could be applied to derive the direct correlation function $c(r)$; (2) a method for the more accurate incorporation of correlation effect could be worked out to extend applicability to higher $\Gamma$-s, where the system is in an intermediate state between liquid and solid phases; (3) application of the current model to anharmonic confinement potentials, like hard-wall traps.

\vspace{5mm}
\begin{acknowledgments}
The authors thank Jeff Wrighton's help on numerical calculation in this paper. The authors are grateful for financial support from the Hungarian Office for Research, Development, and Innovation NKFIH grants K-134462 and K-132158.
\end{acknowledgments}

\bibliography{references}

\end{document}